# Optimizing the nnU-Net model for brain tumor (Glioma) segmentation Using a BraTS Sub-Saharan Africa (SSA) dataset

Authors:
Kalu Chukwuemeka Arua chukwuemeka.kalu.251972@unn.edu.ng
Emegoakor Adaobi Chiazor adaobiemegoakor@gmail.com
Fortune Okafor sirnero001@gmail.com
Okoh Augustine Uchenna okoh.augustineuchenna@gmail.com
Ukpai Chijioke Kelvin ukpaichijioke@gmail.com
Erere Godsent Onyeugbo erereony@gmail.com

**Abstract**

Medical image segmentation is a critical achievement in modern medical science, developed over decades of research. It allows for the exact delineation of anatomical and pathological features in two- or three-dimensional pictures by utilizing notions like pixel intensity, texture, and anatomical context. With the advent of automated segmentation, physicians and radiologists may now concentrate on diagnosis and treatment planning while intelligent computers perform routine image processing tasks.

This study used the BraTS Sub-Saharan Africa (SSA) dataset, a selected subset of the BraTS dataset that included 60 multimodal MRI cases from patients with glioma. Surprisingly, the (no-new net) nnU-Net model trained on the initial 60 instances performed better than the network trained on an offline-augmented dataset of 360 cases. Hypothetically, the offline augmentations introduced artificial anatomical variances or intensity distributions, reducing generalization. In contrast, the original dataset, when paired with nnU-Net's robust online augmentation procedures, maintained realistic variability and produced better results. The study achieved a Dice score of 0.84 for whole tumor segmentation—slightly below the 0.87–0.93 range reported in studies using advanced methods like multi-scale attention and omni-dimensional convolution [30, 33]. For tumor core segmentation, the score was 0.82, also marginally lower than those from previous works [29, 30]. However, the enhancing tumor segmentation results surpassed findings from Asian studies [29, 46]. Overall, the model's performance was comparable to that reported by [12]. These findings highlight the significance of data quality and proper augmentation approaches in constructing accurate, generalizable medical picture segmentation models, particularly for under-represented locations.

## 1.0 Introduction

Medical image segmentation represents a significant advancement in medical science, carefully refined over decades of research and innovation [12]. What is now celebrated as a groundbreaking radiological technique, originated from the need to enhance patient management and satisfaction by accurately modeling medical images. With the advent of automated segmentation, medical professionals and radiologists are increasingly able to focus on diagnosis and treatment planning, as many routine image processing tasks are now handled by intelligent systems [3, 5].

Segmentation of medical images is the process of partitioning radiological images into distinct parts for enhanced structural visualization [11]. It recognizes the boundaries within two-dimensional or three- dimensional visualization using operational concepts including pixel intensity, texture, and anatomical information. Its purpose in an image is to distinguish the target which are relatively complex in terms of morphologies from the background [11, 16]. This is designed to help medical practitioners understand more about a patient's condition by segmenting organs, tissues, or pathological anomalies in medical images, which is vital for precisely localizing aberrations [6]. Convolutional neural networks (CNN) are not dependent on manual image feature extraction or extensive image preprocessing; hence they offer outstanding feature extraction and expression capabilities [3].

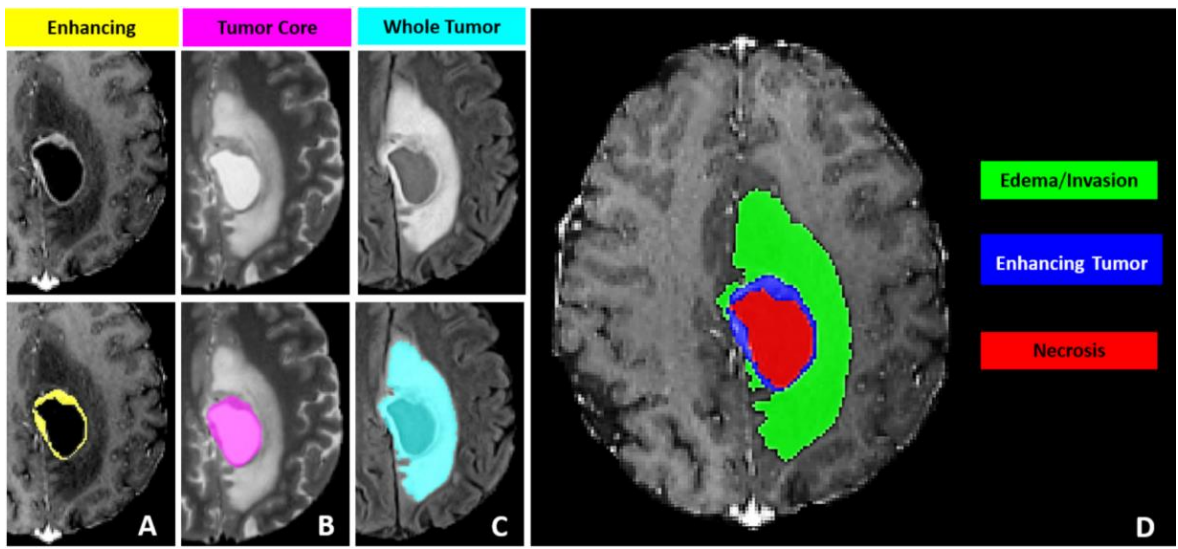

Fig 1 nnU – net applied in segmenting a brain tumor [45]

Image segmentation is progressively replacing traditional image sampling methods. Advances in radiological research have also demonstrated a clear distinction between the manual segmentation and AI-driven models in terms of speed, accuracy, and resource requirements [1, 17]. Manual segmentation is time-consuming, mostly inaccurate, requires more resources, and known to suffer from significant inter-observer variability [10].

Segmentation algorithms are currently widely utilized in most disciplines in medicine, particularly neuro-oncology. It has enhanced the diagnosis of brain tumors, allowing oncologists to detect them earlier and more accurately [4]. Unlike manual segmentation, it is both time and resource intensive. It is well-suited for effective treatment plans and options like surgery, chemotherapy, and radiotherapy, while ensuring patient satisfaction, follow-up and monitoring of patients.

Despite its transformative potential, deep learning in medical image segmentation faces several limitations [7]. One major barrier to widespread clinical adoption is the heterogeneity of acquired imaging data such as variations in contrast, resolution, and signal-to-noise ratios can significantly affect model performance. Deep learning models often struggle to generalize across data from different sources and equipment vendors, leading to inconsistent and suboptimal outcomes [14, 15]. These performance issues are further compounded by intrinsic variability in datasets, the unpredictable nature of optimization processes, and the complexities of selecting appropriate hyper-parameters for both optimization and regularization. Moreover, the architecture of the deep learning models themselves can significantly influence the reliability and accuracy of segmentation results [10, 13].

Achieving an optimal balance between high segmentation accuracy and computational efficiency remains a key objective in the development of deep learning models for medical image analysis. Ensuring that these models are both performant and adaptable to varying hardware constraints is essential for practical clinical deployment [2, 8]. The future of medical diagnostics is promising, driven by ongoing advancements in automated image segmentation. Increasing research efforts are focused on developing algorithms capable of managing the vast variability in medical imaging data, improving generalizability across diverse patient populations, and effectively integrating multimodal imaging inputs to support more accurate and comprehensive diagnoses [9].

This study benchmarks and fine-tunes the self-configuring nnU-Net on the BraTS Sub-Saharan Africa dataset to improve brain tumor segmentation in under-represented locations. By tackling issues in minor lesion detection and dataset diversity, as well as suggesting data augmentation and domain-informed post-processing, the study helps to improve model generalization and clinical application. These advancements are critical for creating egalitarian, high-performing AI algorithms for global medical imaging.

## 2.0     Review of Related Literatures
### 2.1     nnU-Net Model

nnU-Net is not a medical imaging modality itself but a deep learning-based segmentation framework that has significantly advanced the clinical application of automated image segmentation. It employs semantic segmentation techniques and automatically adapts to the characteristics of any given dataset. Built on both 2D and 3D vanilla U-Net architectures, nnU-Net offers a robust and self-configuring framework capable of delivering high accuracy across various medical imaging tasks [20]. A frontier in overcoming long-term medical stagnation in radiological advancement, enabling continual learning in the field of medical imaging. nnU-Net is widely considered as the best performing segmentation for numerous medical applications and includes modules for training, testing, and research model sequencing [2].

This works by supplying training cases and a dataset fingerprint, which configures fresh sets of datasets through its 2D and 3D U-net cascades that function on low resolution and high-resolution images, respectively [20]. nnU-Net is an open-source, self-configuring deep learning system designed for biomedical picture segmentation [4]. Their technology automates the segmented pipeline, establishing any medical dataset, preprocessing, network architecture, training, and post-processing without requiring human intervention. In the context of brain tumors, the appropriateness of nnU-Net for brain segmentation while applying BraTS-specific modification resulted in superior outcomes in 2020 [19], and similarly in the following year with enhanced modalities [22].

A study done by [18] in the lifelong nnU-Net framework for standardization in medical continual learning using various segments modules in evaluation of the prostate, cardiac and hippocampus, proving a better performance in the hippocampus than in the cardiac and prostate which were deteriorated [18]. The brain and spinal cord (collectively known as the Central Nervous System (CNS), the CNS is extremely complex because of its elaborate anatomy [26, 20]. Brain tumor is the abnormal growth of cells in the part of the brain. The exact cause is unknown; however, the risk factors could be enormous. There are 120 types of brain tumor and the gliomas are more common, which are of the grade I and II classification as termed low grade and high grade III and IV [25].

In considering the elevated rate of mortality related to such assessments, prompt identification of brain tumors is essential for effective therapy. For diagnostic reasons, a variety of healthcare visualization techniques, viz. PET, MRI, and CT are utilized [21]. The best technique for assessing soft tissues and the nervous system's function among these is MRI. Though the best technique, U-Net confers a more detailed and better alignment of tumor segmentation from MRI [22].

Before the advent of U-Net algorithms, precision and accuracy of segmented images using conventional imaging techniques was a struggle, especially with the image segmentation, complex structures (noise) and varying textures. Some depended on manual feature extraction methods which was time consuming and less robustly [23, 24].

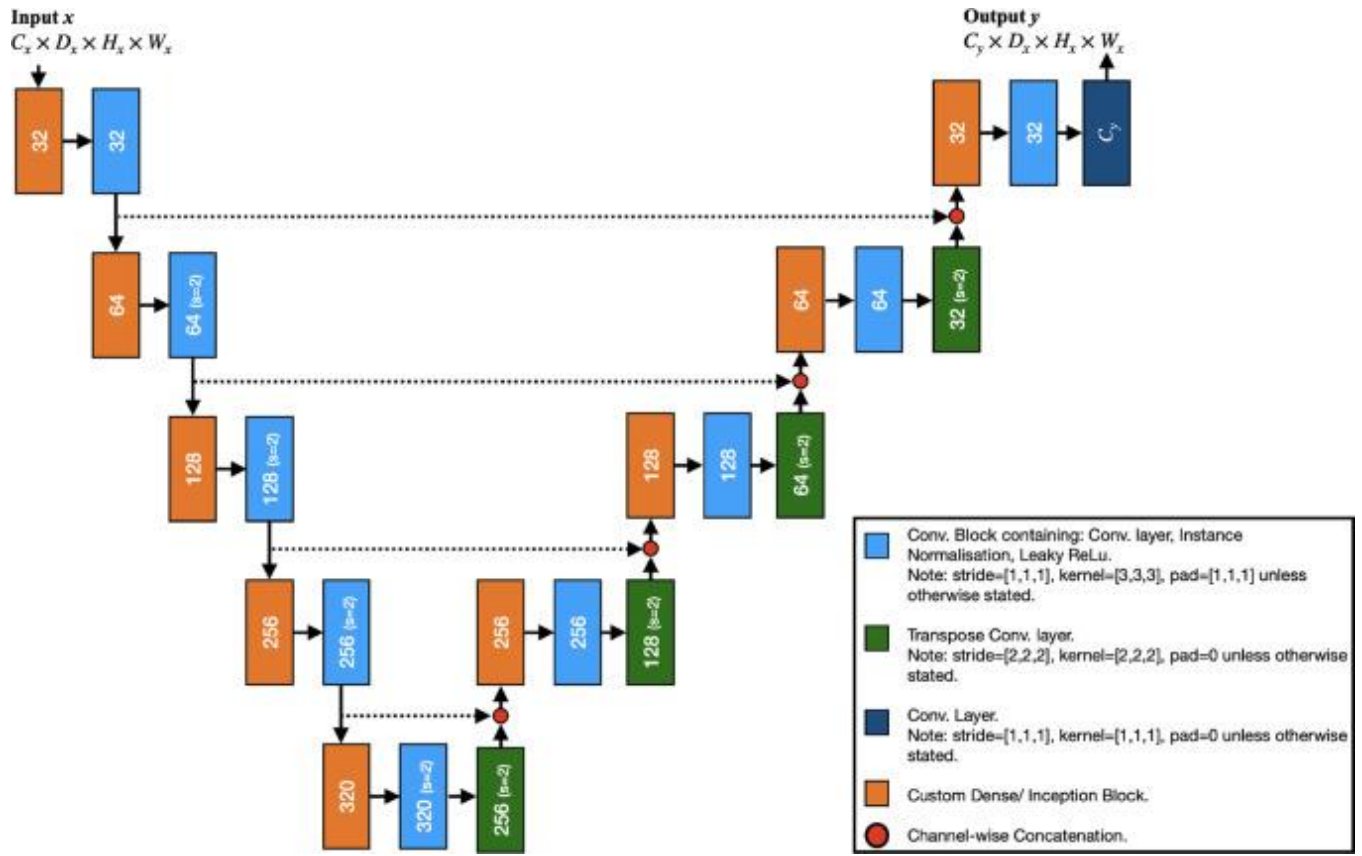

Fig 2 nnU-net model architecture [43].

Precise location of object boundaries and accuracy in data augmentation strategies and dataset architecture organization gives a better performance of U-net as compared to conventional models, U-net is faster and efficient which enables its dependency in processing large images with fast segmentation speeds. Other efficiency is in differentiation of healthy tissues from cancerous ones and delineation of organ structure, crucial for surgical diagnosis [23].

## 2.2 Significance of nnU-Net in Neuro-Oncology
### 2.2.1 Significance of nnU-Net in Tumor Segmentation

*i. State-of-the-Art Performance in Brain Tumor Segmentation*

nnU-Net has consistently demonstrated state-of-the-art performance in brain tumor segmentation, particularly in the BraTS (Brain Tumor Segmentation) challenges. The framework achieved first place in the BraTS 2020 competition with impressive Dice scores of 88.95%, 85.06%, and 82.03% for whole tumor, tumor core, and enhancing tumor, respectively [22], these metrics highlight its ability to accurately delineate tumor regions, which is critical for diagnostic and therapeutic planning.

The success of nnU-Net can be attributed to its ability to leverage multimodal MRI scans, including T1-weighted, T2-weighted, and FLAIR images, to capture complementary information for robust segmentation [29, 30]. Additionally, modifications such as incorporating omni-dimensional dynamic convolution layers and multi-scale attention strategies have further enhanced its performance, particularly in diverse datasets like the BraTS Africa dataset [33].

### ii. Handling Pediatric and Rare Tumor Subtypes

Pediatric brain tumors present unique challenges due to their heterogeneity and the need for precise segmentation of subregions such as enhancing tumor (ET), non-enhancing tumor (NET), cystic components (CC), and peritumoral edema (ED). nnU-Net has shown superior performance compared to other models like DeepMedic, achieving higher Dice scores for these sub-regions [36]. This capability is crucial for pediatric neuro-oncology, where accurate segmentation is essential for treatment planning and monitoring.

### iii. Generalization Across Diverse Datasets

One of the key strengths of nnU-Net is its generalization capability across diverse and unseen datasets. For instance, models trained on multi-institutional pediatric data have shown excellent performance on external validation datasets, including the BraTS-PEDs 2023 dataset [36]. Similarly, nnU-Net has demonstrated robust performance in segmenting brain metastases and gliomas in multi-institutional datasets, further underscoring its versatility [31, 37].

## 2.2.2 Algorithm Efficiency and Computational Considerations

### i. Computational Efficiency

nnU-Net is designed to balance accuracy and computational efficiency, making it suitable for clinical applications where resources may be constrained. By incorporating optimizations such as depthwise-separable convolutions and shuffle attention mechanisms, nnU-Net achieves competitive performance with reduced computational complexity [32]. For example, a modified version of nnU-Net achieved Dice scores of 79.2%, 91.2%, and 84.8% for enhancing tumor, whole tumor, and tumor core, respectively, with only 2.51 million parameters and 55.26 GFLOPS [32].

### ii. Faster Segmentation Times

nnU-Net's efficiency extends to segmentation speed, outperforming traditional algorithms like the mesh growing algorithm (MGA) in terms of time-to-segment. In a comparative study, nnU-Net achieved a mean segmentation time of 1139 seconds compared to MGA's 2851 seconds, making it a more practical choice for neurosurgical settings [28].

### iii. Scalability and Adaptability

The framework's adaptability to different clinical scenarios is another key advantage. For instance, incremental training approaches have been successfully employed to fine-tune nnU-Net for specific institutional settings, ensuring optimal performance while minimizing the need for extensive manual tuning [31].

### 2.2.3 Role in Precision Medicine

#### i. Integration with Clinical Workflows

nnU-Net's ability to integrate with clinical workflows, such as PACS (Picture Archiving and Communication Systems), has been demonstrated in various studies. For example, a PACS-integrated workflow using nnU-Net for brain metastasis segmentation achieved a Dice similarity coefficient (DSC) of 0.85, with sensitivity and specificity of 83% and 92%, respectively [31]. This integration facilitates seamless clinical application, enabling rapid and reproducible tumor segmentation.

#### ii. Reduction in Inter-Observer Variability

Manual segmentation by radiologists is time-consuming and prone to inter-observer variability. nnU-Net has been shown to reduce this variability, achieving segmentation results comparable to expert radiologists while minimizing the need for manual adjustments [35]. This consistency is critical for precision medicine, where reliable and reproducible results are essential.

#### iii. Longitudinal Glioma Segmentation

nnU-Net has also been evaluated for longitudinal glioma segmentation, demonstrating its ability to segment phenotypic regions such as necrosis, contrast enhancement, and edema across pre- and post-treatment MRI scans [34]. This capability is vital for monitoring disease progression and treatment response, enabling personalized treatment strategies.

Table 1: Performance Comparison of nnU-Net across Different Brain Tumor Segmentation Tasks

| Tumor Region | Dice Score Range | Key Enhancements | Citation |
| --- | --- | --- | --- |
| Whole Tumor (WT) | 0.87–0.93 | Multi-scale attention, omni-dimensional convolutions | (Mistry et al., 2024; Guo et al., 2023) |
| Tumor Core (TC) | 0.83–0.89 | Transformer integration, deep supervision | (Guo et al., 2023; Fang & Huang, 2024) |
| Enhancing Tumor (ET) | 0.77–0.86 | Attention mechanisms, domain knowledge infusion | (Fang & Huang, 2024; Kotowski et al., 2020) |

## 3.0 Methodology
### 3.1 Dataset Description

In 3D medical image segmentation, the objective is to assign a semantic label $y_i \in C$ to each voxel $x_i \in \Re$, where $C = \{0,1,...C-1\}$ is the set of class labels. Given an input volume $X = \{x_1, x_2,...x_n\}$, the segmentation model $f_\theta(X)$ predicts the class probabilities for each voxel using a function parameterized by weight $\theta$. The prediction output is a tensor $\hat{Y} \in \Re^{n \times C}$, where each voxel's prediction is a C-dimensional probability vector (22).

**Convolutional Neural Networks (CNN)**

The core operation in U-Net and nnU-Net architectures is the 3D convolution, which computes features maps by convolving a 3D kernel $K \in \Re^{d \times d \times d}$, with a 3D input volume $X \in \Re^{H \times W \times D}$.

$$Y(i,j,k) = \sum_u^d \sum_v^d \sum_w^d K(u,v,w).X(i+u, j+v, k+w) \qquad 1$$

This operation is repeated across multiple layers to learn increasingly abstract representations of the input data.

This study utilized the BraTS Sub-Saharan Africa (SSA) dataset, a curated subset of the BraTS dataset designed to support brain tumor segmentation research in underrepresented regions. The dataset comprises 60 multimodal MRI cases from patients diagnosed with glioma, a type of brain tumor arising from glial cells [12].

From [12], each case includes the following four MRI modalities, each offering complementary anatomical and pathological information:
 i. T1-weighted (T1): Provides high-resolution structural information.
 ii. Gadolinium-enhanced T1-weighted (T1Gd): Highlights areas with a compromised blood-brain barrier, typically enhancing tumor regions (ET).
 iii. T2-weighted (T2): Sensitive to fluid content and useful in identifying edema and non-enhancing tumor components.
 iv. Fluid-Attenuated Inversion Recovery (FLAIR): Suppresses CSF signals and highlights peritumoral edema or infiltrative tumor margins.

Ground truth segmentations, annotated according to the BraTS protocol [38], include:
 i. Enhancing Tumor (ET): Regions with contrast uptake in T1Gd.
 ii. Tumor Core (TC): Combination of ET, necrotic, and non-enhancing tumor.
 iii. Surrounding Non-enhancing FLAIR Hyperintensity (SNFH): Represents infiltrative tumor or edema.

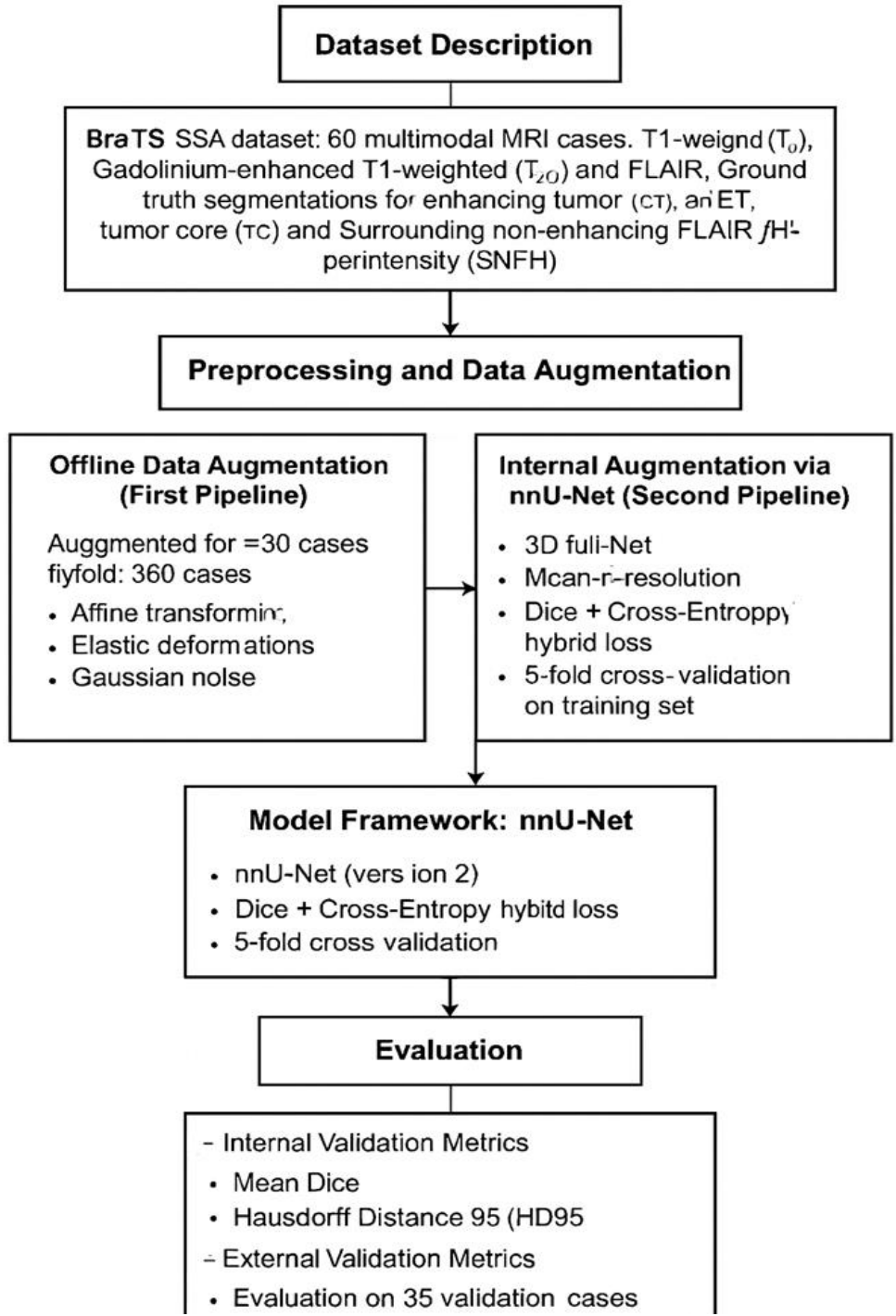

Fig 3 Stages of image segmentation using the Brats dataset

**3.2.    Preprocessing and Data Augmentation**
**3.2.1   Offline Data Augmentation (First Pipeline)**

Data augmentation acts as a regularizer and helps deep neural networks generalize better on unseen data [44]. However, improper augmentation can lead to unrealistic samples and poor model generalization.

Offline augmentation was performed using TorchIO [44], a Python library tailored for 3D medical image preprocessing and augmentation.

To compensate for the small dataset size, each of the 60 original cases was augmented five times, yielding a final dataset of 360 samples. The transformations aimed to increase diversity and reduce overfitting:
  i. Random Affine Transformations: Simulate changes in patient orientation.
  ii. Elastic Deformations: Mimic realistic anatomical variability.
  iii. Gaussian Noise Addition: Introduce scanner-based or physiological noise.

### 3.2.2  Internal Augmentation via nnU-Net (Second Pipeline)
On-the-fly augmentation retains original sample distribution and avoids dataset drift while improving robustness to common image acquisition variations [22].
The second pipeline relied exclusively on nnU-Net's internal augmentation, applied on-the-fly during training. This includes:
  i. Random Rotation and Scaling
  ii. Gamma Correction (brightness variation)
  iii. Axis-wise Mirroring

These transformations are dynamically adjusted during training to retain anatomical plausibility and dataset-specific intensity statistics.

### 3.3.3  Model Framework: nnU-Net
The study adopted the nnU-Net framework [22], known for its self-configuring capability to adapt to different biomedical segmentation tasks without manual architecture tuning. The nnU-Net's automated configuration allows it to outperform many custom-tuned models across a range of datasets due to its strong inductive biases and robust optimization routines [22, 41].

### 3.3.4  Network Configuration
  i. Architecture: 3D Full-Resolution U-Net
  ii. Loss Function: Dice + Cross-Entropy Hybrid Loss — Dice loss handles class imbalance; cross-entropy penalizes misclassification.
  iii. Cross-Validation: 5-fold for original data
  iv. Optimizer: Stochastic Gradient Descent (SGD) with Nesterov momentum — accelerates convergence by anticipating gradient direction.

v. Learning Rate Schedule: Polynomial decay, which gradually reduces learning rate to stabilize learning.
vi. Batch Size, Patch Size, and Spacing: Automatically computed by nnU-Net based on GPU memory and data properties.

**The Dice Loss**

The Dice similarity coefficient (DSC) is a measure of overlap between the predicted segmentation $\hat{Y}$ and the ground truth $Y$.

$$DSC = \frac{2\sum_{i=1}^{n} \hat{y}_i y_i}{\sum_{i=1}^{n} \hat{y}_i + \sum_{i=1}^{n} y_i} \quad\quad 2$$

The Dice loss is hence:

$$L_{Dice} = 1 - DSC \quad\quad 3$$

This loss is effective in handling class imbalance, which is common in medical images, where tumor regions are much smaller than the background.

Cross-Entropy Loss: The voxel-wise classification, the categorical cross-entropy loss is defined as:

$$L_{CE} = -\sum_{i=1}^{n} \sum_{c=1}^{C} y_{i,c} \log(\hat{y}_{i,c}) \quad\quad 4$$

Where,

$y_{i,c} \in \{0,1\}$ is the ground truth for voxel $i$ and class $c$,

$\hat{y}_{i,c} \in [0,1]$ is the predicted probability.

The final hybrid loss used is:

$$L = L_{Dice} + L_{CE} \quad\quad 5$$

### 3.3 Training Strategy and Experimental Setup

While larger datasets are generally beneficial for deep learning [40], the realism and representativeness of the augmented data are equally important for generalization. Two experiments were carried out using Lightning AI infrastructure:

Table 2 Image segmentation approach in the two experiments

| Experiment 1: Augmented Dataset Training | Experiment 2: Original Dataset Training |
| --- | --- |
| Input Data: 360 cases (60 original + 300 offline-augmented) | Input Data: 60 original cases only |
| Training Duration: 100 epochs | Training Duration: 85 epochs |
| Validation: 1-fold internal validation | Validation: 5-fold cross-validation |
| Test-Time Augmentation: None | Augmentation: Real-time by nnU-Net |

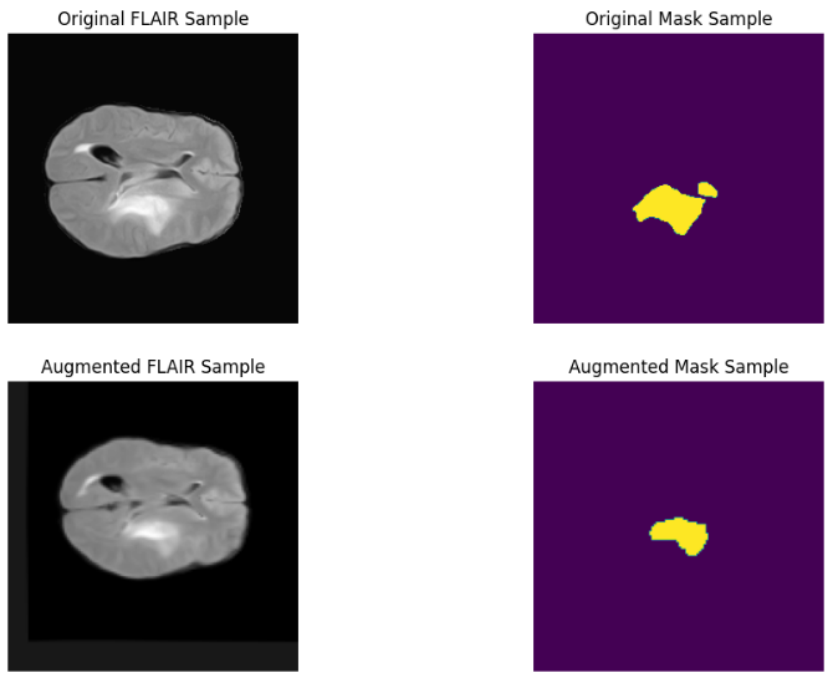

Fig 4 Tumor identification in image segmentation using offline augmentation

### 3.4    Evaluation Metrics

**1. Internal Validation Metrics**

  i.  Mean Dice Coefficient: Measures the overlap between predicted and reference segmentations.
  ii. Hausdorff Distance (HD95): Captures the largest surface distance between segmentations after outlier exclusion (95th percentile), important for assessing spatial accuracy.

**2. External Validation Metrics (Synapse Evaluation Platform)**

Evaluation metrics on 35 held-out cases included:
- Lesion-Wise Dice Scores for ET, TC, WT
- Normalized Surface Dice (NSD) at 0.5 mm and 1.0 mm tolerances
- Aggregate Dice Scores

These metrics are commonly adopted in BraTS evaluations and emphasize both region accuracy and surface boundary precision [39, 42].

## 4.0. Summary of Results

Below are the results obtained using the nnU-Net model:

Table 3 summary of the metric report on the image segmentation

| Metric Type | Experiment 1 (Augmented) | Experiment 2 (Original) |
|---|---|---|
| LesionWise_Dice_ET | 0.767656859085484 | 0.8361266520156425 |
| LesionWise_Dice_TC | 0.778996627986313 | 0.8220899763278429 |
| LesionWise_Dice_WT | 0.877215464426858 | 0.8849396733899775 |
| LesionWise_NSD_0.5_ET | 0.36270723806322197 | 0.5684872360350618 |
| LesionWise_NSD_0.5_TC | 0.34441544768892374 | 0.5012520316401545 |
| LesionWise_NSD_0.5_WT | 0.3627265363487104 | 0.5030484851932235 |
| LesionWise_NSD_1.0_ET | 0.7176171067815279 | 0.8349065898863797 |
| LesionWise_NSD_1.0_TC | 0.6687534433384057 | 0.7579181432757582 |
| LesionWise_NSD_1.0_WT | 0.7243807024808828 | 0.7904024089030094 |
| Dice_ET : | 0.8419531190940647 | 0.8941782634203005 |
| Dice_TC | 0.8573187327245607 | 0.8941504552379538 |
| Dice_WT | 0.9155201025470284 | 0.9246915075937181 |
| NSD_0.5_ET | 0.3934959954877287 | 0.6043145184600957 |
| NSD_0.5_TC | 0.3705346628537408 | 0.5330446229036137 |
| NSD_0.5_WT | 0.3772370763677439 | 0.5268466285145458 |
| NSD_1.0_ET | 0.781787311043553 | 0.8909529700467725 |
| NSD_1.0_TC | 0.7248685697483873 | 0.8150056449542454 |
| NSD_1.0_WT | 0.7537131168365606 | 0.8256568495936597 |
| Cases Evaluated | 35 | 35 |

## 5.0 Discussion

Contrary to expectations, the model trained on the original 60 cases significantly outperformed the model trained on the augmented 360-case dataset. We hypothesize that the offline augmentations, though varied, may have introduced unrealistic anatomical variations or intensity distributions not representative of true patient scans, thereby hindering generalization. In contrast, the original dataset, combined with nnU-Net's robust online augmentation, preserved more meaningful variability and yielded better performance.

The result in our study thus suggests that offline augmentations may introduce anatomical or statistical inconsistencies that degrade model generalization. Moreso, the internal augmentation by nnU-Net preserved anatomical integrity and true clinical variability, enhancing learning effectiveness.

This aligns with prior findings that dataset quality and augmentation fidelity often surpass raw quantity in determining model performance in medical imaging tasks [45].

The study obtained a Dice score of 0.84 for whole tumor segmentation, which is slightly lower than scores reported in other studies (ranging from 0.87 to 0.93) which employed enhancements such as multi-scale attention and omni-dimensional convolution [30, 33]. Similarly, the Dice score for tumor core segmentation was 0.82, which was slightly lower than those reported in previous study settings [29, 30]. Interestingly, the score for enhancing tumor segmentation outperformed results from Asian studies [29, 46]. Overall, the model performed similarly to the results published by [12]. This further explains how performance can be affected by intrinsic variability in datasets, especially with different sources, the unpredictable nature of optimization processes, and the complexities of selecting appropriate hyper-parameters.

## 6.0 Conclusion

This study emphasizes the importance of data quality and proper augmentation procedures when constructing robust and generalizable medical picture segmentation models. The higher performance of nnU-Net on the original BraTs-SSA dataset demonstrates the limitations of indiscriminate offline augmentation, particularly when it generates artificial variances. Emphasizing realistic data variability and adaptable frameworks like nnU-Net is critical for increasing automated segmentation, particularly in resource-limited and under-represented healthcare contexts.